\documentclass[12pt,oneside, a4paper]{article}

\ifx\pdfoutput\undefined
\usepackage[dvips,bookmarks=false]{hyperref}	
\else
\usepackage{hyperref}	
\fi
\hypersetup{colorlinks,bookmarksopen,bookmarksnumbered,citecolor=blue,
linkcolor=blue,pdfstartview=FitH,urlcolor=blue}

\usepackage{graphicx}
\usepackage{amssymb}
\usepackage{cite}
\usepackage{bm}
\usepackage{amsmath,amsthm}

\newcommand{\capdef}{}
\newcommand{\mycaption}[2][\capdef]{\renewcommand{\capdef}{#2}%
       \caption[#1]{{\footnotesize #2}}}

\newcommand{\be}{\begin{equation}}
\newcommand{\ee}{\end{equation}}

\newcommand{\dmq}{\ensuremath{\Delta m^2_{31}}}

\begin{document}

\begin{titlepage}

\begin{center}

\vspace*{2cm}
{\Large\bf Determination of the neutrino mass ordering by\\[2mm] 
combining PINGU and Daya Bay II}
\vspace{1cm}

\renewcommand{\thefootnote}{\fnsymbol{footnote}}
{\bf Mattias Blennow}\footnote[1]{emb AT kth.se}, 
{\bf Thomas Schwetz}\footnote[2]{schwetz AT mpi-hd.mpg.de},
\vspace{5mm}

$^*$ {\it Department of Theoretical Physics,
School of Engineering Sciences, KTH Royal Institute of Technology,
AlbaNova University Center, 106 91 Stockholm, Sweden}

$^\dagger$ {\it%
{Max-Planck-Institut f\"ur Kernphysik, Saupfercheckweg 1, 69117 Heidelberg, Germany}}

\vspace{8mm} 

\abstract{The relatively large measured value of $\theta_{13}$ has
  opened various possibilities to determine the neutrino mass ordering,
  among them using PINGU, the low-energy extension of the IceCube
  neutrino telescope, to observe matter effects in atmospheric
  neutrinos, or a high statistics measurement of the neutrino energy
  spectrum at a reactor neutrino experiment with a baseline of around
  60~km, such as the Daya Bay~II project. In this work we point out a
  synergy between these two approaches based on the fact that when data
  are analysed with the wrong neutrino mass ordering the best fit occurs at different values of $|\dmq|$ for PINGU and Daya
  Bay~II. Hence, the wrong mass ordering can be excluded by a mismatch
  of the values inferred for $|\dmq|$, thanks to the excellent
  accuracy for $|\dmq|$ of both experiments. We perform numerical
  studies of PINGU and Daya Bay~II sensitivities and show that the
  synergy effect may lead to a high significance determination of the
  mass ordering even in situations where the individual experiments
  obtain only poor sensitivity.}

\end{center}
\end{titlepage}

\renewcommand{\thefootnote}{\arabic{footnote}}
\setcounter{footnote}{0}

\setcounter{page}{2}

\section{Introduction}

Experiments with atmospheric neutrinos and reactor antineutrinos have
played a crucial role for our current understanding of neutrino
oscillations. The first evidence for neutrino oscillations was
obtained by the observation of the zenith angle dependent
disappearance of atmospheric muon neutrinos in
SuperKamiokande~\cite{Fukuda:1998mi}. The KamLAND reactor experiment
provided an independent confirmation of solar neutrino oscillations
\cite{Eguchi:2002dm} and obtained strong evidence for the 
spectral distortion as predicted by
oscillations~\cite{Araki:2004mb}. Various reactor experiments also played a crucial role in the determination of the last unknown mixing
angle $\theta_{13}$~\cite{Apollonio:2002gd, An:2012eh, Ahn:2012nd,
  Abe:2012tg}. Combining those with other data on oscillations
\cite{Ahmad:2002jz, Adamson:2008zt, Abe:2011sj} we now have a good
picture of the neutrino mass pattern and the leptonic mixing
matrix, see~\cite{GonzalezGarcia:2012sz} for a global fit. One of the
big remaining questions is the type of the neutrino mass ordering,
which can be ``normal'' or ``inverted'', depending on the sign of the
neutrino mass-squared difference \dmq\ responsible for the above
mentioned muon neutrino disappearance. Indeed, there is a chance that
atmospheric and/or reactor neutrinos again may play a crucial role in
answering this question.

In atmospheric neutrino experiments, the determination of the mass ordering is
based on the matter effect~\cite{Wolfenstein:1977ue, Barger:1980tf,
  Mikheev:1986gs} in \dmq\ driven oscillations, see~\cite{Blennow:2013rca} for a recent review.  In case of normal
ordering ($\dmq > 0$) the MSW resonance will occur for neutrinos,
whereas for the inverted ordering ($\dmq < 0$) it will happen for
antineutrinos. Atmospheric neutrinos with baseline lengths up to the
full diameter of the earth provide an interesting opportunity to study this
effect. One possibility is to invoke a magnetic field to separate
neutrino- and antineutrino-induced muons. This approach is pursued by
the ICal@INO experiment~\cite{INO}, for recent sensitivity studies
see~\cite{Blennow:2012gj,Ghosh:2012px}. In the absence of a magnetic
field no discrimination between neutrino- and antineutrino-induced
events is possible on an event-by-event basis, and therefore the
effect of changing the neutrino mass ordering is strongly diluted.
However, a non-zero net effect remains, since neutrinos and
antineutrinos do not contribute equally to the total event sample due to different interaction cross sections.  In
huge detectors with several Mt yr exposures those subtle effects are
observable and provide sensitivity to the neutrino mass ordering.
This possibility is discussed in the context of neutrino telescopes such as
IceCube and ANTARES/KM3NET under the acronyms PINGU (Precision IceCube
Next-Generation Upgrade)~\cite{Koskinen:2011zz} and ORCA (Oscillation
Research with Cosmics in the Abyss)~\cite{orca}, respectively, or for
huge underground detectors~\cite{Abe:2011ts, Autiero:2007zj}.

A very different method to determine the neutrino mass ordering has
been pointed out in~\cite{Petcov:2001sy}. In a reactor experiment
close to the first oscillation maximum of $\Delta m^2_{21}$ (at a
baseline around 60~km), oscillations driven by $\Delta m^2_{31}$ and
$\theta_{13}$ manifest themselves as small wiggles in the energy
spectrum. The effect of the neutrino mass ordering enters via a subtle
interference effect in the $\bar\nu_e\to\bar\nu_e$ survival
probability between oscillations due to $\dmq$ and $\Delta
m^2_{21}$. The basic observation is that the amplitudes of
oscillations with $\Delta m^2_{31}$ and $\Delta m^2_{32}$ are
different, due to the large but non-maximal value of
$\theta_{12}$. Furthermore, for normal ordering $|\Delta m^2_{31}| >
|\Delta m^2_{32}|$, whereas for inverted ordering $|\Delta m^2_{31}| <
|\Delta m^2_{32}|$. Hence, by finding out whether the larger or the
smaller frequency in the energy spectrum has the larger or smaller
amplitude one can determine the mass ordering. The possibilities for
exploring this effect are currently investigated in the context of the
Daya Bay II~\cite{DB2-YWang,DB2-WWang} and the RENO50~\cite{reno50}
projects.

Both of these methods to determine the neutrino mass ordering are
experimentally challenging. They require huge statistics as well as
excellent control over energy reconstruction and for atmospheric
neutrinos also directional reconstruction. Currently it is not fully
established whether the necessary requirements can be met and it might
happen that those experiments achieve only modest sensitivity to the
mass ordering. In the following we point out an interesting way to
identify the mass ordering with high significance by combining data from
an atmospheric and a reactor neutrino experiment, even if the individual
sensitivities are low. To be specific we will consider the PINGU and
Daya Bay~II setups. 

The main observation is the following: when a fit to the data is
performed with the wrong mass ordering, the best fit point will be
located at a value $|\dmq|_\text{wrong}$, which is different to the
one in the correct mass ordering. Moreover, the value of
$|\dmq|_\text{wrong}$ is different in the atmospheric and reactor
neutrino experiments. It turns out that the difference between
$|\dmq|_\text{wrong}^\text{PINGU}$ and $|\dmq|_\text{wrong}^\text{Daya
  Bay II}$ typically is larger than the uncertainties with which it
will be determined by those experiments. Hence, by the mismatch of
the values for $|\dmq|$ determined by the two experiments one may be
able to exclude the wrong mass ordering, even if the individual
$\chi^2$ differences between the fit with the correct and the wrong
ordering are small.

A similar effect has been discussed in~\cite{Nunokawa:2005nx,
  Minakata:2006gq}, where the comparison of $\nu_\mu$ and $\nu_e$
disappearance experiments is used to determine the mass ordering, see
also~\cite{deGouvea:2005hk}. The main idea is similar to our proposal
of combining PINGU and Daya Bay~II, however there are important
differences. In~\cite{Nunokawa:2005nx} the focus is on disappearance
experiments measuring $\dmq$ at the first oscillation maximum. The
formulas derived in that reference for the effective mass-squared
differences $\Delta m^2_{ee}$ and $\Delta m^2_{\mu\mu}$ are based on
this configuration. In the experiments we are considering here the
oscillation physics are somewhat more involved. For Daya Bay~II the
wrong $|\dmq|$ is determined by fitting the tiny wiggles on the energy
spectrum with the wrong mass ordering. Some analytic considerations on
the value of $|\dmq|_\text{wrong}^\text{Daya Bay
  II}$ can be found in~\cite{Qian:2012xh}. Also, the oscillation
physics in PINGU are more complicated than for a long-baseline $\nu_\mu$
disappearance experiment. In PINGU a superposition of
the $\nu_e \to \nu_\mu$ and $\nu_\mu \to \nu_\mu$ channels is
observed, while also summing over neutrinos and antineutrinos. Furthermore,
a wide range of energies and baselines is sampled, including resonant
matter effects. This makes it more difficult to understand the
location of $|\dmq|_\text{wrong}^\text{PINGU}$ analytically. We will
determine the locations of the $\chi^2$ minimum with the wrong mass ordering numerically, and
indeed we do find deviations from the formulas derived in
\cite{Nunokawa:2005nx} comparable to the precision of $|\dmq|$. In fact, we will see that the best fit with the wrong ordering differs from that in the correct one also when $\Delta m_{21}^2 = 0$.

The combination of Daya Bay~II data with an independent determination
of $|\dmq|$ from MINOS or T2K has been considered in, e.g.,~\cite{Qian:2012xh, Ge:2012wj, Li:2013zyd}, where a modest
improvement of the sensitivity (depending on the assumed accuracy) is
obtained. A combination of PINGU data with a measurement of $|\dmq|$
from T2K has been performed in Ref.~\cite{Winter:2013ema}, finding
only a marginal improved mass ordering sensitivity by this
combination. Indeed, both PINGU as well as Daya Bay~II will both
obtain unprecedented precision in the determination of $|\dmq|$ and
therefore they offer the most promising combination to explore this
effect for the determination of the mass ordering.

The outline of the paper is as follows. We start discussing the
sensitivity of PINGU and Daya Bay~II individually in sections
\ref{sec:pingu} and \ref{sec:DB2}, respectively, where also the details of
our numerical analyses are given. In section~\ref{sec:comb} we show
how an excellent sensitivity to the mass ordering can be obtained by
the combination of PINGU and Daya Bay~II, even when the individual
experiments do not achieve a high sensitivity. We conclude in
section~\ref{sec:conclusions}.

\section{PINGU}
\label{sec:pingu}

PINGU~\cite{Koskinen:2011zz} is a low-energy extension of the IceCube
neutrino telescope at the south pole, obtained by installing
additional strings with digital optical modules~(DOMs) within the
existing DeepCore detector~\cite{Collaboration:2011ym}. This would
allow the observation of atmospheric neutrinos with a fiducial volume
of several Mt and a threshold of a few GeV.\footnote{The possibility
  to use such a detector as target for an artificial neutrino beam has
  been considered in \cite{Fargion:2010vb, Tang:2011wn,
    Brunner:2013lua}.} It has been pointed out in
\cite{Akhmedov:2012ah} that, with the large value of $\theta_{13}$
that has recently been established, such a configuration has the
potential to determine the neutrino mass ordering, see also
\cite{Mena:2008rh, FernandezMartinez:2010am}. As discussed in detail
in~\cite{Akhmedov:2012ah}, the difference between the mass orderings
is visible as a characteristic pattern in the plane of neutrino energy
and direction. A crucial detector requirement is therefore a good
ability to reconstruct neutrino energy and direction, see also
\cite{Indumathi:2004kd, Petcov:2005rv, Samanta:2006sj} in the context
of a magnetized detector. Recently the sensitivity of a PINGU-like
configuration has been studied by a number of authors
\cite{Akhmedov:2012ah, Agarwalla:2012uj, Franco:2013in,
  Ribordy:2013xea, Winter:2013ema}. Given the importance of the issue
we will also provide a further independent study of the PINGU setup,
considering various options for the achieved resolutions and show the
impact on the sensitivity. For the sake of definiteness we will
concentrate on the capabilities of PINGU; qualitative features are
expected to be similar for ORCA.

\subsection{Event rate calculation}

In this work we consider only muon-like events
and assume an effective detector mass for PINGU as obtained by recent
simulations of the IceCube collaboration: we adopt the effective volume
times density of ice as function of neutrino energy as shown on slide
3 of~\cite{cowen-talk} by the curve labeled ``Triggered Effective
Volume, $R=100$~m''. This curve can be approximated by the expression
\begin{equation}\label{eq:Veff}
  \rho_{\rm ice} V_{\rm eff}(E_\nu) \approx 2.4  \, 
  \log\left(\frac{E_\nu}{1 \, {\rm GeV}} - 2 \right)^{0.87} \, {\rm Mt} \,. 
\end{equation}
The threshold is around 3 GeV and the effective mass rises to about
4~Mt at 10~GeV and 7~Mt at 35~GeV. We consider the initial neutrino
flavours $\nu_\mu, \bar\nu_\mu, \nu_e, \bar\nu_e$ and fold the fluxes
with the oscillation probabilities $P_{\nu_\mu \to\nu_\mu},
P_{\bar\nu_\mu \to \bar\nu_\mu}, P_{\nu_e \to\nu_\mu}, P_{\bar\nu_e
  \to \bar\nu_\mu}$, respectively. The probabilities are obtained by
numerically solving the evolution equations through a realistic Earth
density profile~\cite{Dziewonski:1981xy}.

The ability to reconstruct neutrino energy and direction is crucial to
resolve the signatures induced by changing the neutrino mass
ordering. The actual performance of the PINGU detector is under active
investigation. Here we adopt various assumptions to show the goals
which have to be achieved in order to obtain a relevant sensitivity.
We consider two different approaches: muon reconstruction or neutrino
reconstruction, which we describe in the following.


{\bf Muon reconstruction.} Here we (conservatively) assume that only
the muon can be reconstructed and no information from the hadron
shower is used (this approach has been adopted in~\cite{Franco:2013in}). 
We implement this by an integration of the event rates
based on a Monte Carlo sample of neutrino events\footnote{We thank
  Anselmo Meregaglia for providing us his Monte Carlo event sample to
  perform this study.}. Neutrino nadir angles are sampled in
$\cos\theta_\nu$ between 0 and 1 in steps of $\Delta\cos\theta_\nu =
0.02$. For given $\cos\theta_\nu$ a random neutrino energy is drawn
according to the atmospheric neutrino flux~\cite{Honda:1995hz},
weighted by the effective detector mass as a function of neutrino
energy and the oscillation probability. 

A muon is then generated with the GENIE event generator
\cite{Andreopoulos:2009rq} assuming a water target, returning the muon
energy $E_\mu$ and the angle $\alpha$ between the muon and the
neutrino. The muon energy is smeared assuming a Gaussian detector
resolution $\sigma_{E_\mu}$ and distributed accordingly over the bins
in $E_\mu$. We use 20 bins in $E_\mu$ logarithmically spaced between
1~GeV and 40~GeV (note that the actual threshold is set by the
effective volume function in Eq.~\ref{eq:Veff}). We use logarithmic
bins in order to have finer binning in the low energy region, where
the main information on the mass ordering is obtained. We also assume
a Gaussian detector resolution $\sigma_{\theta_\mu}$ on the muon
direction. This is implemented by drawing a random angle
$\delta\alpha$ from a Gaussian distribution with width
$\sigma_{\theta_\mu}$ which we then add to the angle $\alpha$. We then
distribute the event in bins in the muon nadir angle using the given
neutrino nadir angle and the angle $\alpha + \delta\alpha$ between
neutrino and muon, assuming a flat distribution of the muon azimuthal
angle with respect to the neutrino direction. We use 20 bins in
$\cos\theta_\mu$ between 0.1 and 1.

Monte Carlo events are generated for an exposure of 100~years and scaled
down to the desired exposure time in order to predict the expected
event numbers per bin. Even for perfect reconstruction of the muon
energy and direction, a sizable smearing of the oscillation
probability happens due to the kinematical relations between neutrino
and muon energies/directions. We show some representative values of
the muon reconstruction abilities to illustrate the impact on the
sensitivity. Apart from the perfect muon detection with $\sigma_{E_\mu} =
0$ and $\sigma_{\theta_\mu} = 0$, we consider also the situation of
$\sigma_{E_\mu} = 15\%$ and $\sigma_{\theta_\mu} = 10^\circ$.


{\bf Neutrino reconstruction.} Most likely the assumption that no
information from the hadron shower can be obtained is too
pessimistic. It may be possible to reconstruct the neutrino energy and
direction directly by using information from the muon track as well as
the hadron shower\footnote{In~\cite{Ribordy:2013xea} also the
  reconstruction of the $y$-distribution is considered, which may in
  principle be used to discriminate neutrino from antineutrino events
  on a statistical basis. We do not follow this strategy
  here. According to~\cite{Ribordy:2013xea} using the $y$-distribution
  leads to an increase of significance between about 10\% (no
  degeneracy, worse resolutions) up to 42\% (worse case degeneracy,
  best resolution). See also~\cite{Ghosh:2013mga}.}. Estimates on the
reconstruction accuracies for neutrino properties are given in
\cite{Akhmedov:2012ah}. We adopt the same representative values as
used there. The neutrino energy and angle reconstruction resolutions
are assumed to be Gaussian with the widths
\begin{equation}\label{eq:pingu-res}
\sigma_{E_\nu} = A + B E_\nu \,,\qquad
\sigma_{\theta_\nu} = C
\sqrt{\frac{1\,{\rm GeV}}{E_\nu}} \,,
\end{equation}
respectively. For the energy resolution we consider the two examples
($A = 2 \,{\rm GeV}, B = 0$) and ($A = 0, B = 0.2$) and for the
angular resolution we take the two representative values $C = 0.5$ and
$C = 1$~rad. For $C = 0.5$ this corresponding to about $13^\circ$
($9^\circ$) at $E_\nu = 5$~GeV (10~GeV). For the calculation of the event
rates we follow closely~\cite{Petcov:2005rv}, where more details can
be found.  The neutrino fluxes~\cite{Honda:2004yz} are folded with the
cross section~\cite{Paschos:2001np}, the neutrino energy dependent
effective detector mass, the oscillation probabilities as well as with
the reconstruction resolutions described above. The same binning as
for the muon data is used, except that now it is understood in terms
of reconstructed neutrino quantities.

\subsection{Statistical analysis and systematic uncertainties}
\label{sec:chisq}

We investigate the potential to determine the mass ordering by
performing a statistical analysis based on a $\chi^2$-function.  We
consider the two-dimensional event distribution in either muon or neutrino
nadir angle and energy.  Let us denote the number of events in
bin $jk$ by $R_{jk}(\boldsymbol{x})$, where $\boldsymbol{x}$ is a vector of 
the oscillation parameters.  We calculate ``data'' by adopting ``true
values'' $\boldsymbol{x}^\mathrm{true}$ for the oscillation
parameters: $D_{jk} = R_{jk}(\boldsymbol{x}^\mathrm{true})$.  In the
theoretical prediction we take several sources of
systematic errors into account by introducing 11 pull variables $\boldsymbol{\xi} =
(\xi_1,\ldots,\xi_{11})$:
\begin{equation}\label{eq:pred}
T_{jk}(\boldsymbol{x},\boldsymbol{\xi} ) = 
R_{jk}(\boldsymbol{x}) 
\left( 1 + \sum_{l=1}^{11} \xi_l \, \pi^l_{jk} \right)\,,
\end{equation}
with appropriately defined ``couplings'' $\pi^l_{jk}$.  In the
systematic error treatment we follow closely the description given in
the appendix of~\cite{GonzalezGarcia:2004wg}, where also a definition
of the $\pi^l_{jk}$ can be found. The
systematic effects included in our analysis are listed in
Tab.~\ref{tab:systematics}. We consider a fully correlated overall
normalization error of 20\% from various sources such as uncertainties in the
atmospheric neutrino fluxes, the cross sections, the fiducial detector
mass, or efficiencies. 
Furthermore, we take into account an uncertainty in the
neutrino/antineutrino ratio (including fluxes as well as cross
sections) and an error on the ratio of $e$-like to $\mu$-like fluxes.
In addition to these normalization errors we also allow for
uncertainties in the shape of the neutrino fluxes by introducing
a linear tilt in the nadir angle as well as in energy, uncorrelated between the
four fluxes of $\nu_e, \bar\nu_e, \nu_\mu, \bar\nu_\mu$.

\begin{table}
\centering
\begin{tabular}{clr}
\hline\hline
index $l$ & systematic effect & value \\
\hline
1 & overall normalization & 20\% \\
2 & $\nu/\bar\nu$ ratio & 5\% \\
3 & $\nu_\mu/\nu_e$ ratio of fluxes & 5\% \\
$4-7$ & $\cos\theta_{\nu}$ dependence of fluxes & 5\%\\
$8-11$ & energy dependence of fluxes & 5\%\\
\hline\hline
\end{tabular}
\mycaption{Systematic uncertainties included in our PINGU analysis.
\label{tab:systematics}}
\end{table} 

We adopt a $\chi^2$-definition based on Poisson statistics:
\begin{equation}\label{eq:chisq}
\Delta\chi^2(\boldsymbol{x}) = 
\min_{\boldsymbol{\xi}} 
\left[
2 
\sum_{j=1}^{N^\mathrm{bin}_E} 
\sum_{k=1}^{N^\mathrm{bin}_\theta} 
\left(
T_{jk}(\boldsymbol{x},\boldsymbol{\xi} ) - D_{jk} + 
D_{jk} \ln \frac{D_{jk}}{T_{jk}(\boldsymbol{x},\boldsymbol{\xi})}
\right)
+
\sum_{l=1}^{11} \xi_l^2
\right] \,.
\end{equation}
The dependence on $\boldsymbol{x}^\mathrm{true}$ is implicit via
$D_{jk}$. In the cases of interrest (PINGU and Daya Bay~II) event
numbers in most of the bins are large and in this case the first term
of the right-hand side of eq.~\ref{eq:chisq} becomes equivalent to the
standard Gaussian $\chi^2$ definition. The term including the pull
parameters $\xi_l$ assumes that systematic uncertainties behave like
Gaussian errors. The sensitivity based on pure
statistics (without including the effect of systematic uncertainties)
is obtained by fixing the $\xi_l$ to zero.  We call eq.~\ref{eq:chisq}
``$\Delta\chi^2$'' because for $\boldsymbol{x} =
\boldsymbol{x}^\mathrm{true}$ eq.~\ref{eq:chisq} is zero, by
definition. In order to quantify the sensitivity to the mass ordering
we consider $\Delta \chi^2(\boldsymbol{x})$ where the sign of $\dmq$
in $\boldsymbol{x}$ is taken opposite to the one in
$\boldsymbol{x}^\mathrm{true}$. Whenever we quote $\Delta \chi^2$
values for the mass ordering sensitivity we always minimize $\Delta
\chi^2(\boldsymbol{x})$ with respect to $|\dmq|$, without including
any external information on this parameter. We will show the impact of
either fixing $\theta_{13}$ and $\theta_{23}$ or minimizing with
respect to them. When minimizing over them we include external
information on the mixing angles via adding a term $\chi^2_{\rm
  prior}$ to eq.~\ref{eq:chisq}, assuming a 5\% error on
$\sin^2(2\theta_{13})$ as well as a 15\% error on
$\sin^2(2\theta_{23})$.
If not stated otherwise we use the following true values:
\begin{equation}\label{eq:true-values}
\begin{split}
  &|\dmq| = 2.4\cdot 10^{-3}~{\rm eV}^2 \,,\qquad
  \Delta m^2_{21} = 7.59\cdot 10^{-5}~{\rm eV}^2 \,,\\
  &\sin^2 2\theta_{13} = 0.09 \,,\quad
  \sin^2 2\theta_{23} = 1 \,, \quad
  \sin^2\theta_{12} = 0.302 \,, \quad
  \delta = 0 \,.
\end{split}
\end{equation}

The precise statistical meaning of sensitivity statements based on
eq.~\ref{eq:chisq} is non-trivial. A detailed discussion of those
issues is given in~\cite{BCHS}, see also~\cite{Qian:2012zn,
  Ciuffoli:2013rza}. One can define a test statistic which under
certain conditions follows a Gaussian distribution with mean given by
$\Delta \chi^2$ from eq.~\ref{eq:chisq} and standard deviation
$2\sqrt{\Delta \chi^2}$~\cite{BCHS, Qian:2012zn, Franco:2013in}. In
this paper we will follow the traditional habit of saying that ``the
sensitivity to the mass ordering is $n\sigma$'' with $n = \sqrt{\Delta
  \chi^2}$. It can be shown~\cite{BCHS} that under the above mentioned
Gaussian assumption the interpretation of the statement is the
following: if $n = \sqrt{\Delta \chi^2}$ then the experiment will
exclude the wrong mass ordering at $n\sigma$ with a probability of
roughly 50\%. We have checked~\cite{BCHS} that for PINGU and even more
for Daya Bay~II typically those assumptions are approximately
fulfilled, in reasonable agreement with the results from
\cite{Franco:2013in} for PINGU and \cite{Qian:2012xh} for Daya Bay~II.

\subsection{PINGU sensitivity to the mass ordering}
\label{sec:results}

Let us first consider the effect of energy and directional
resolutions, systematic uncertainties, and parameter degeneracies on the sensitivity to the mass ordering. For
this study we will adopt the approximation $\Delta m^2_{21} = 0$,
motivated by the fact that we consider only neutrino energies above
several GeV. In that range effects of $\Delta m^2_{21}$ are expected
to be small. In this limit the mixing angle $\theta_{12}$ as well as
the CP phase $\delta$ become unphysical, and oscillation
probabilities depend only on the three parameters $\theta_{13}$,
$\theta_{23}$ and $\Delta m^2$, where $\Delta m^2 \equiv \dmq = \Delta
m^2_{32}$ for $\Delta m^2_{21} = 0$.  This approximation is necessary
for practical purposes, in order to make the numerical analysis
including systematic uncertainties as well as the marginalization over
$\theta_{13}$ and $\theta_{23}$ feasible. We will discuss the
effect of the CP phase $\delta$ without adopting any approximation on
the three-flavour oscillations later in this work.

\begin{figure}[t]
\begin{center}
\includegraphics[width=0.49\textwidth]{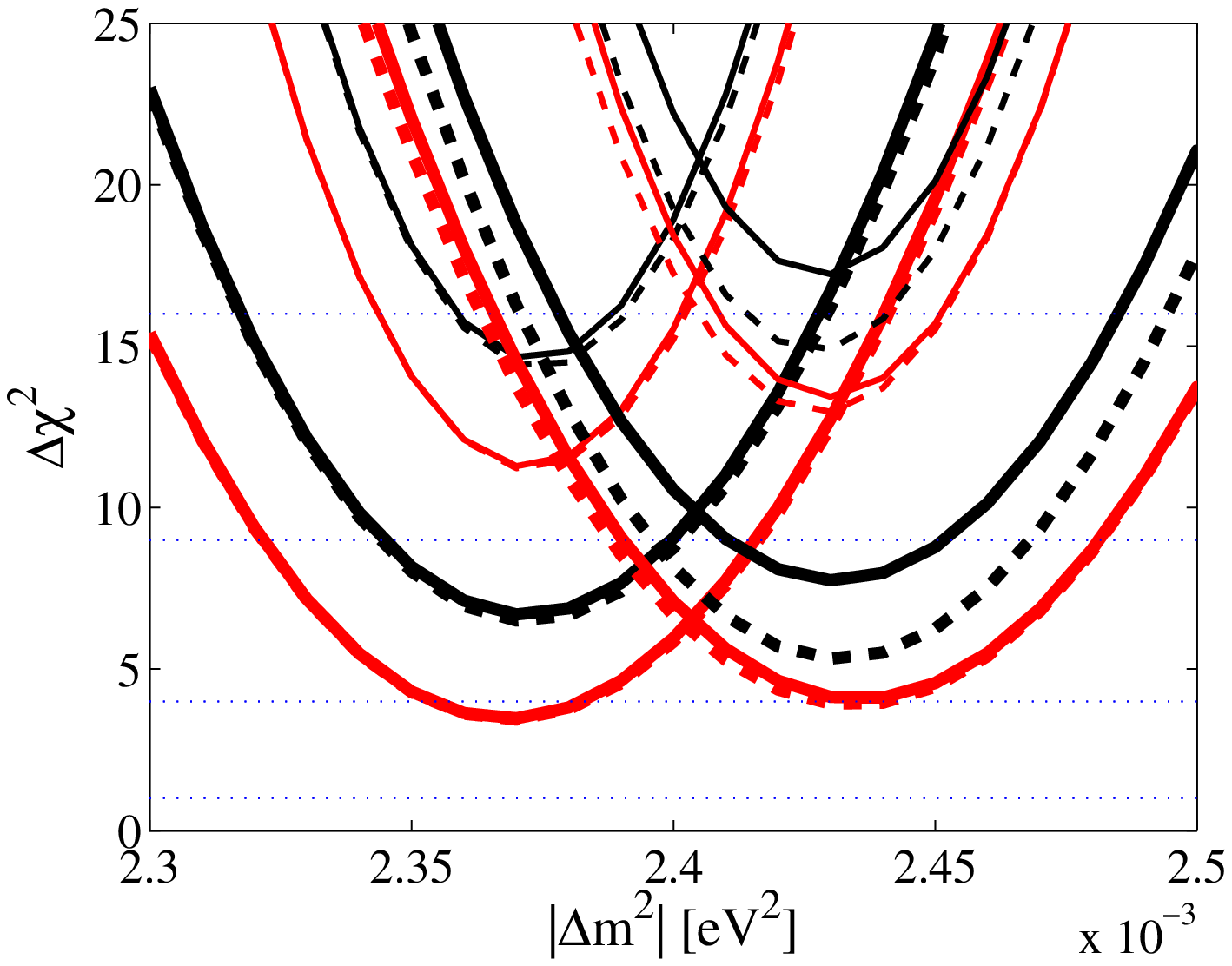}
\includegraphics[width=0.49\textwidth]{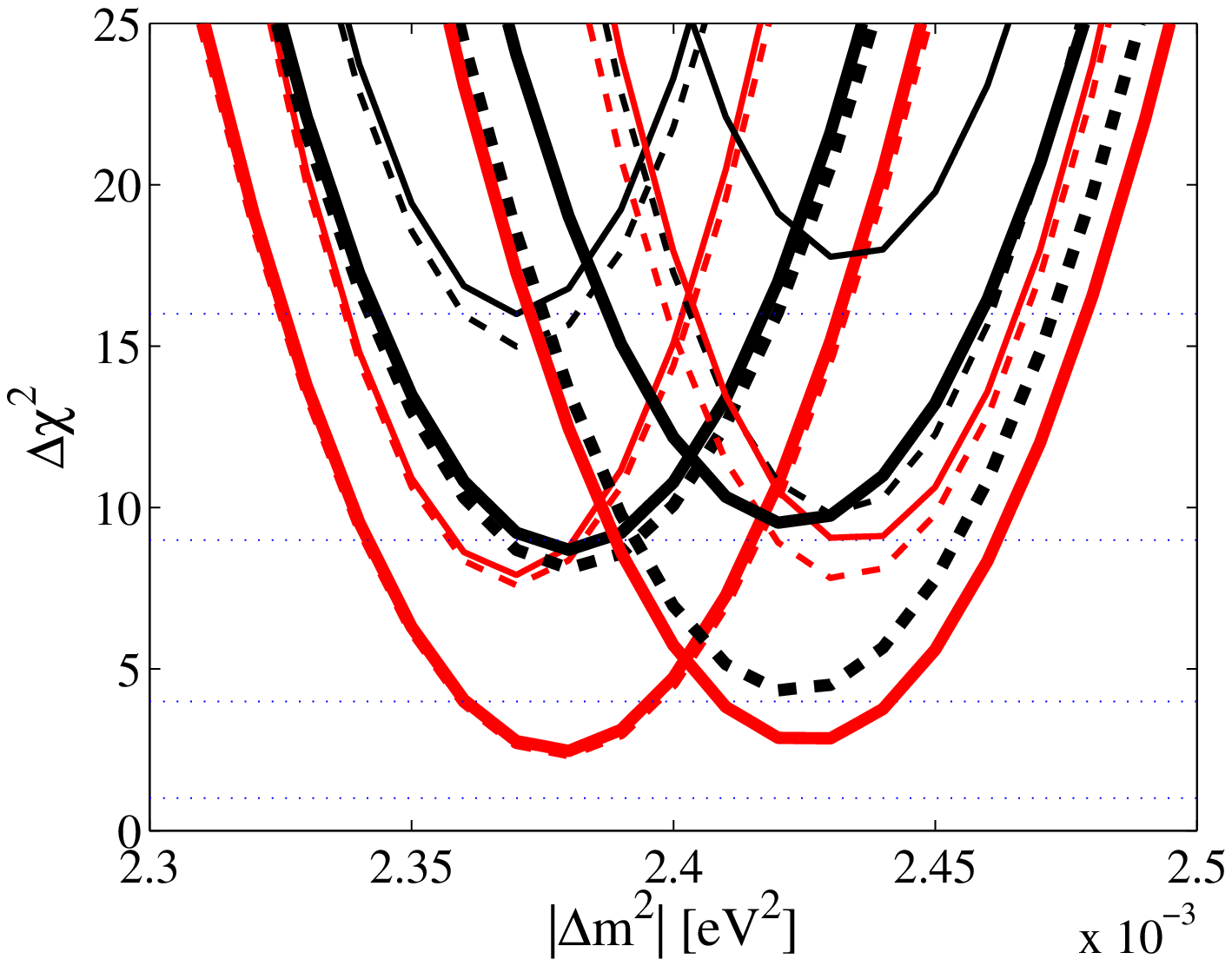}
\end{center}
\mycaption{The expected PINGU 3-year $\Delta\chi^2$ in the wrong ordering as a function of $|\Delta m^2|$ (in the approximation $\Delta m^2_{21} = 0$) for the simulations involving the muon parameters (left) and for the simulations using the reconstructed neutrino parameters (right). The solid (dashed) curves correspond to simulations where the oscillation parameters were fixed to their input values (allowed to vary with a prior penalizing too large deviations from them), the black (red) curves correspond to simulations without (with) systematic errors included, and the different thickness of the curves signify different assumptions on the energy and angular resolutions. For the left panel, the resolutions were assumed to be $\sigma_{E_\mu}/E_\mu = 0.15$ and $\sigma_{\theta_\mu} = 10^\circ$ for the thick curves, while perfect energy and angular resolution on the muon was assumed for the thin curves. For the right panel, the resolution on the reconstructed neutrino parameters were assumed to be $\sigma_{E_\nu} = 2$~GeV ($0.2E_\nu$) and $\sigma_{\theta_\nu} = \sqrt{(1~{\rm GeV})/E_\nu}$ ($0.5\sqrt{(1~{\rm GeV})/E_\nu}$) for the thick (thin) curves. In both panels, the right (left) curves correspond to the normal (inverted) ordering fit to a true inverted (normal) ordering.}\label{fig:dm2}
\end{figure}

In Fig.~\ref{fig:dm2}, we show the expected $\Delta\chi^2$ as a
function of $|\Delta m^2|$ of the wrong mass ordering after 3 years of
data and how it depends on our assumptions regarding systematic
errors, marginalization of parameters, and experimental
resolutions. The set of curves at smaller (larger) values of $|\dmq|$
correspond to a true normal (inverted) ordering. We note that in all
cases, systematic errors will have a significant impact on the final
sensitivity of the experiment. We find that the uncertainty in the oscillation
parameters $\theta_{23}$ and $\theta_{13}$ are important only for true
inverted ordering. However, once both effects have been
taken into account, the effects of the other are diminished. In
particular, the effect of allowing parameters to vary freely is
significantly reduced once the systematic errors have been introduced.
Despite the differences introduced by taking these effects into account,
the most crucial assumption is the assumption on the energy and
angular resolutions, with the results varying with several $\sigma$
depending on the assumptions made (compare thin versus thick
curves). It should also be clear from the figure that whether we make
assumptions on the measurements of the muon parameters (left panel) or
those of the reconstructed neutrino parameters (right panel), the
results are relatively similar and seem to exhibit similar
characteristic when it comes to the importance of systematic errors
and incomplete knowledge of the neutrino parameters.

\begin{figure}
\begin{center}
\includegraphics[width=0.49\textwidth]{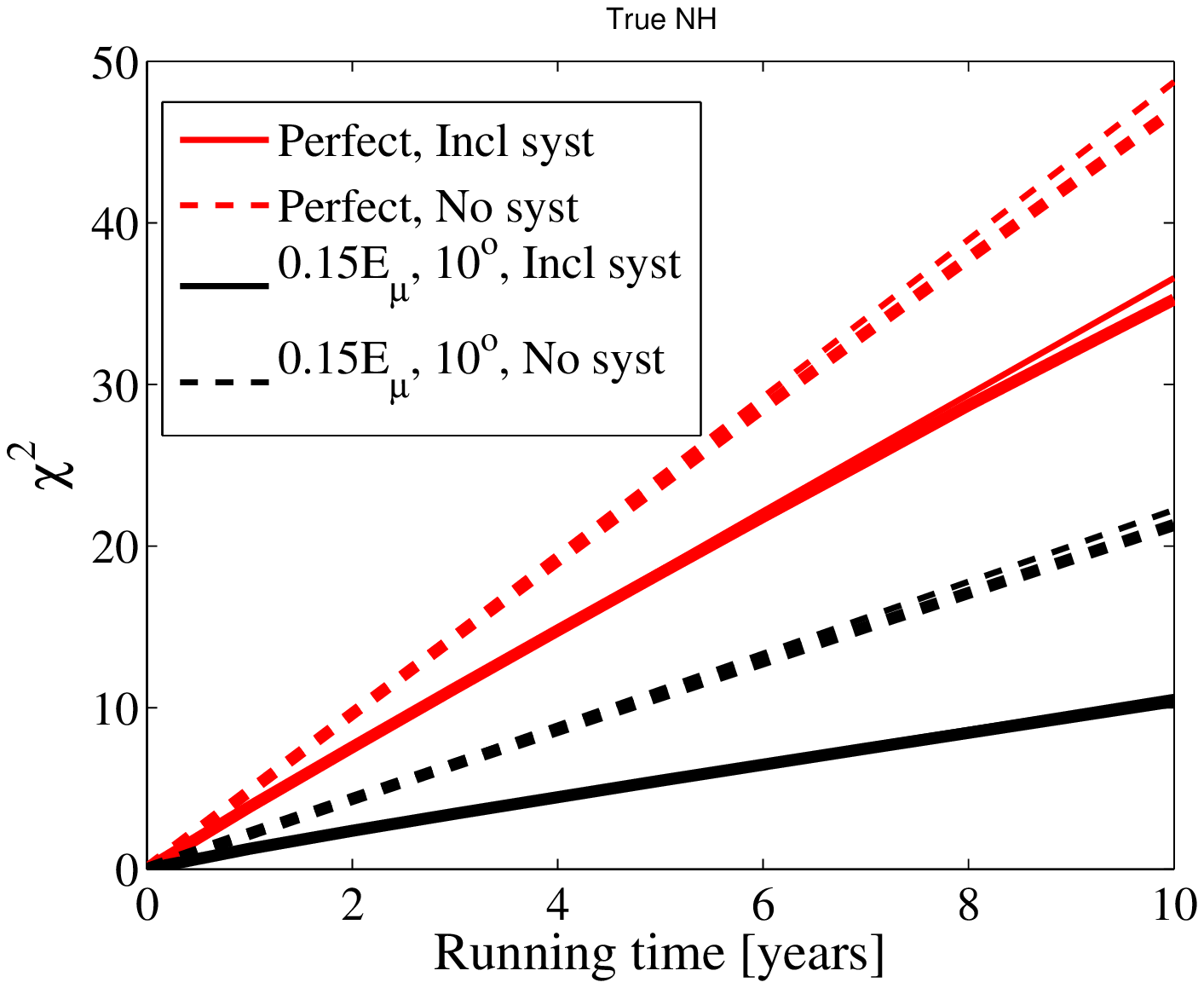}
\includegraphics[width=0.49\textwidth]{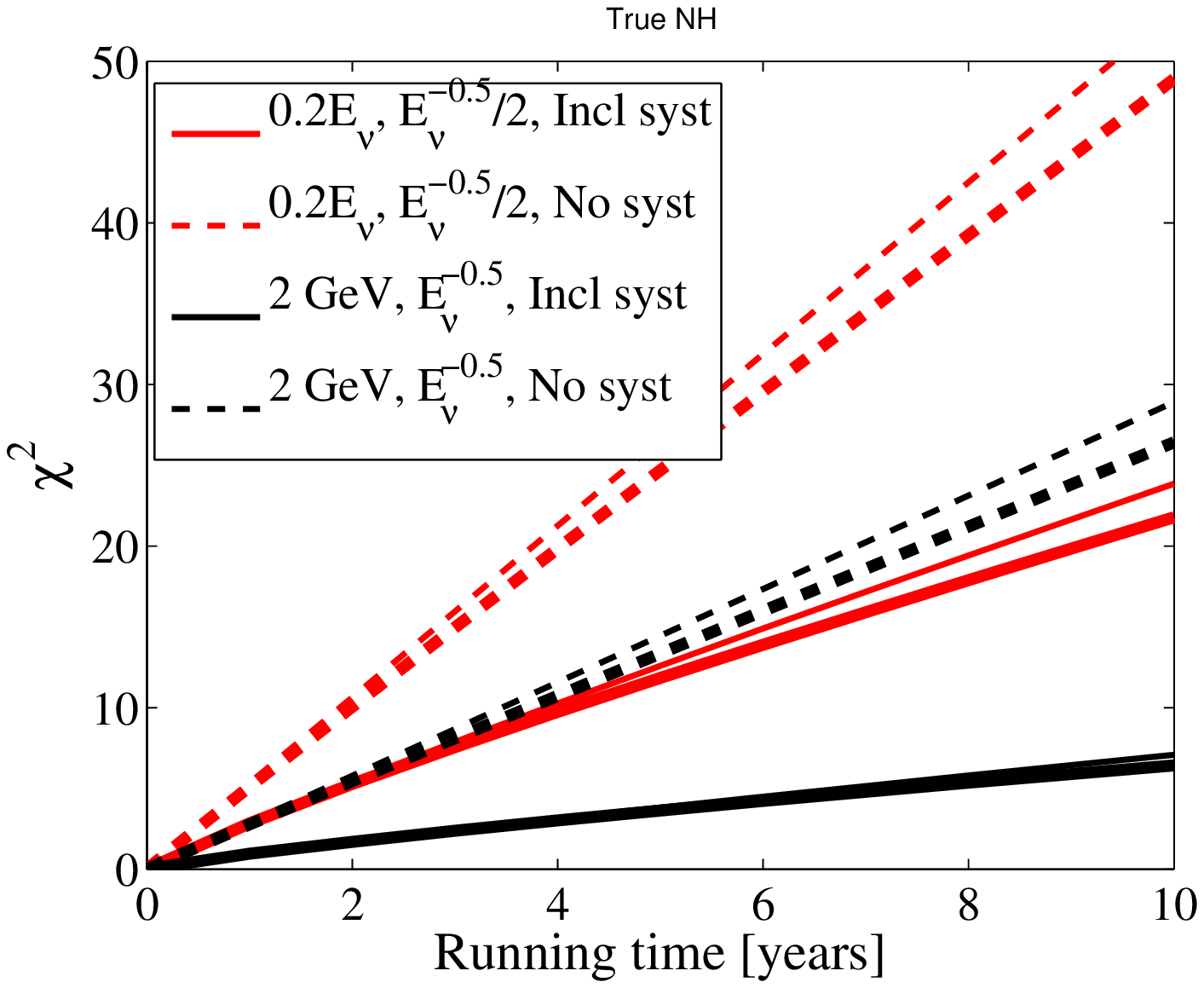}
\end{center}
\mycaption{Time evolution of the PINGU mass ordering discovery potential for our different experimental assumptions on the energy and angular resolutions of the detector (see labeling in the figure). The left panel corresponds to using the muon parameters and the right panel to using the reconstructed neutrino parameters. The thin (thick) lines correspond to fixing (marginalizing) over $\theta_{23}$ and $\theta_{13}$. All curves are marginalized with respect to $|\dmq|$ and we have assumed normal ordering to be true.}\label{fig:timeline}
\end{figure}

In Fig.~\ref{fig:timeline}, we show the sensitivity of PINGU to the
mass ordering under the assumption that the normal ordering is the
true one as a function of the PINGU running time.  We confirm previous
results that the energy and angular resolutions play a major role in
the mass ordering determination (e.g.,~\cite{Akhmedov:2012ah,
  Winter:2013ema}), as does the inclusion of systematic
errors. Comparing red versus black curves with the same style shows
the impact of the resolutions, whereas dashed versus solid curves with
the same color shows the impact of systematics. From the comparison of
thin versus thick curves we find that minimizing with respect to $\theta_{13}$
and $\theta_{23}$ has only a marginal impact on the sensitivity. Note,
however, that in all cases we do minimize with respect to $|\Delta m^2|$, which
is very important. By comparing the left and the right panel we find
that a very similar behaviour is obtained by our two methods of taking
into account reconstruction, either using only the muon or working
with reconstructed neutrino parameters. In summary, for our setups,
the ones with the best resolution and without systematic errors would
have a sensitivity of around seven sigma after ten years of running;
the sensitivity deteriorates to around three sigma if systematics
are taken into account and only the more conservative assumptions on
the resolutions can be satisfied.

\begin{figure}
\begin{center}
\includegraphics[width=0.49\textwidth]{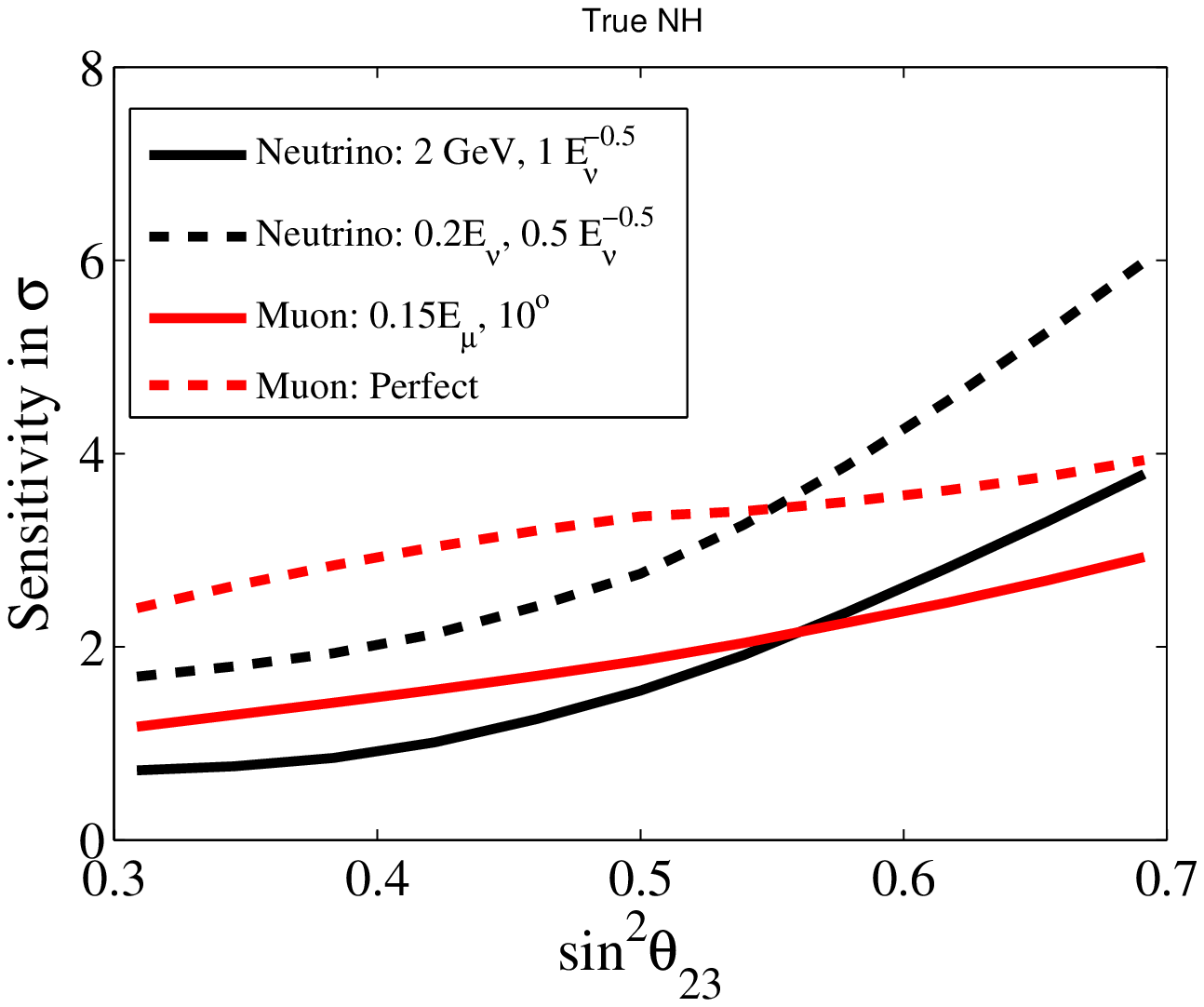}
\includegraphics[width=0.49\textwidth]{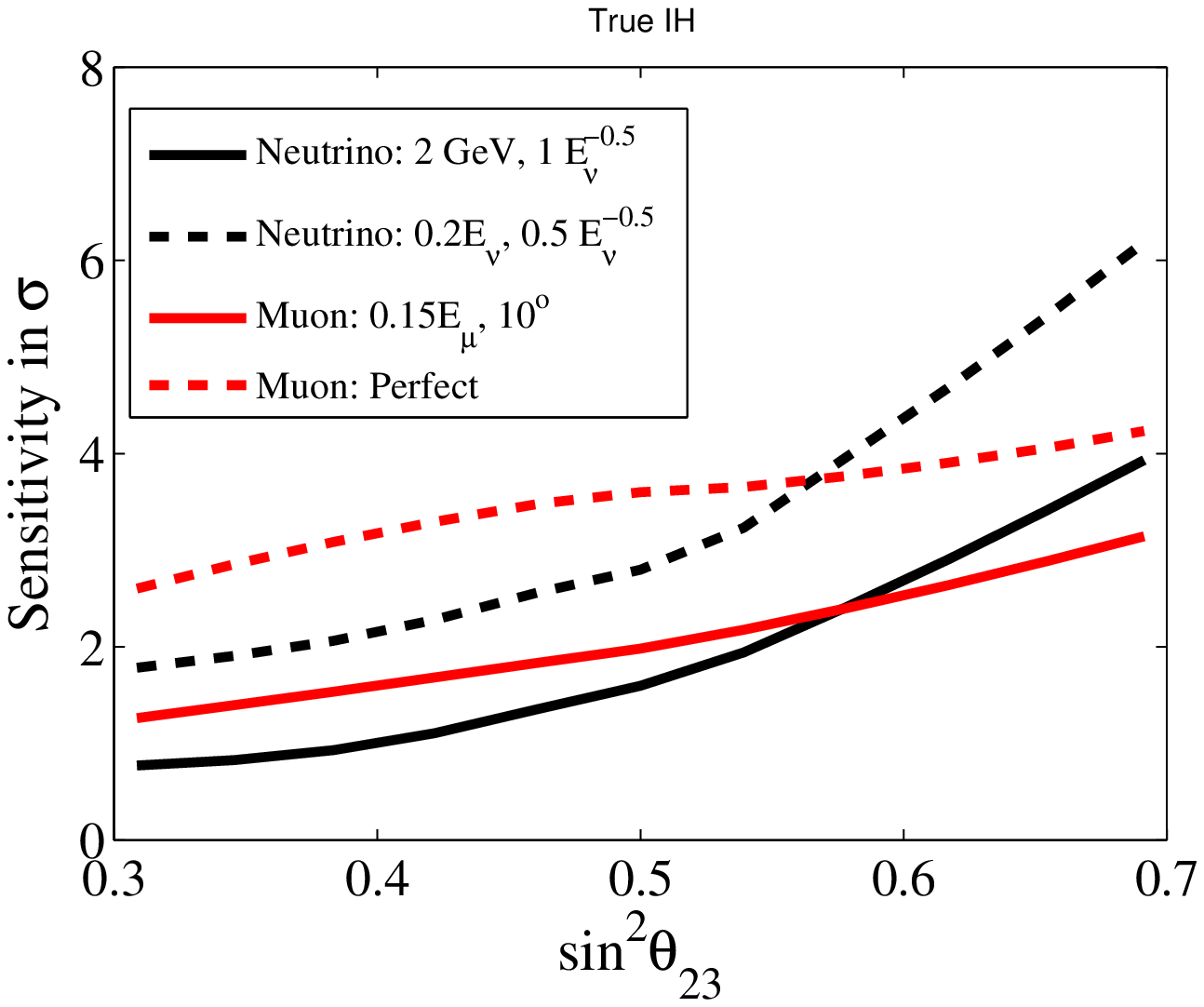}
\end{center}
\mycaption{Dependence of the PINGU mass ordering sensitivity after
  3 years on the true value of $\theta_{23}$. The
  sensitivity is defined as $\sqrt{\Delta\chi^2}$.  The black (red)
  curves correspond to our setups studying the reconstructed neutrino
  (muon) parameters. The true neutrino mass ordering assumed to be
  normal (inverted) for the left (right) panel, and we include
  systematic effects as well as marginalization over the neutrino
  oscillation parameters. Dashed versus solid curves correspond to
  different assumptions on the resolutions as given in the
  legend.}\label{fig:octantPINGU}
\end{figure}

It is well known that the sensitivity of atmospheric neutrino
experiments to the neutrino mass ordering is quite dependent on the
true value of the lepton mixing angle $\theta_{23}$. While there is so
far no evidence for deviation of $\theta_{23}$ from the maximal mixing
value $\pi/4$, there are some hints at a level below two sigma that
non-maximal values are preferred~\cite{GonzalezGarcia:2012sz}.  In
Fig.~\ref{fig:octantPINGU}, we show the dependence of the PINGU
sensitivity on the true value of $\sin^2\theta_{23}$. We find that,
while the general trend of better sensitivity for $\theta_{23}$ in the
second octant observed in other studies is present, the size of the
effect is strongly dependent not only on the assumptions we make on
the detector resolutions, but also on whether we treat the problem
using the muon or reconstructed neutrino parameters, with the
reconstructed neutrino parameters gaining significantly in sensitivity
for large values of $\theta_{23}$, while the muon parameter analysis
shows a more modest gain.  On a quantitative level, the sensitivity to
the neutrino ordering as given by the reconstructed neutrino
parameters changes by more than three sigma within the currently
allowed range for $\theta_{23}$, from a low sensitivity of slightly
below two sigma at $\sin^2\theta_{23} = 0.34$ to almost five sigma at
$\sin^2\theta_{23} = 0.67$ (for the more optimistic assumption on the
resolutions). For the analysis using the muon parameters, the
difference in sensitivity between the extreme values of $\theta_{23}$
is only around one and a half sigma.
We interpret this behaviour such that the smearing
affects $\theta_{23}$ induced signatures in a slightly different way,
depending on whether neutrino or muon quantities are used to describe
the resolution. Since both approaches adopted here are approximate and
should be replaced by realistic reconstruction algorithms based on
Monte Carlo studies including detailed experimental information we do
not investigate this behaviour further. We just conclude that also the
$\theta_{23}$ dependence of the sensitivity seems to depend crucially
on the actual energy and direction reconstruction abilities of the
detector. Let us mention also that by comparing the left and right panels we find
that results are rather independent on whether the true ordering is
normal or inverted.

\begin{figure}[t]
\begin{center}
\includegraphics[width=0.37\textwidth]{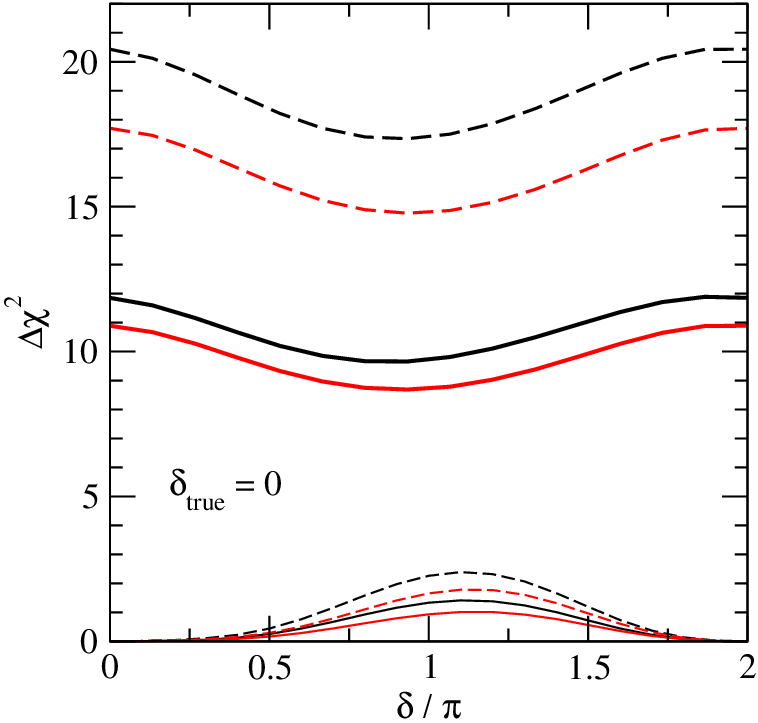}
\includegraphics[width=0.37\textwidth]{{chisq-del-0.5}.eps}\\
\includegraphics[width=0.37\textwidth]{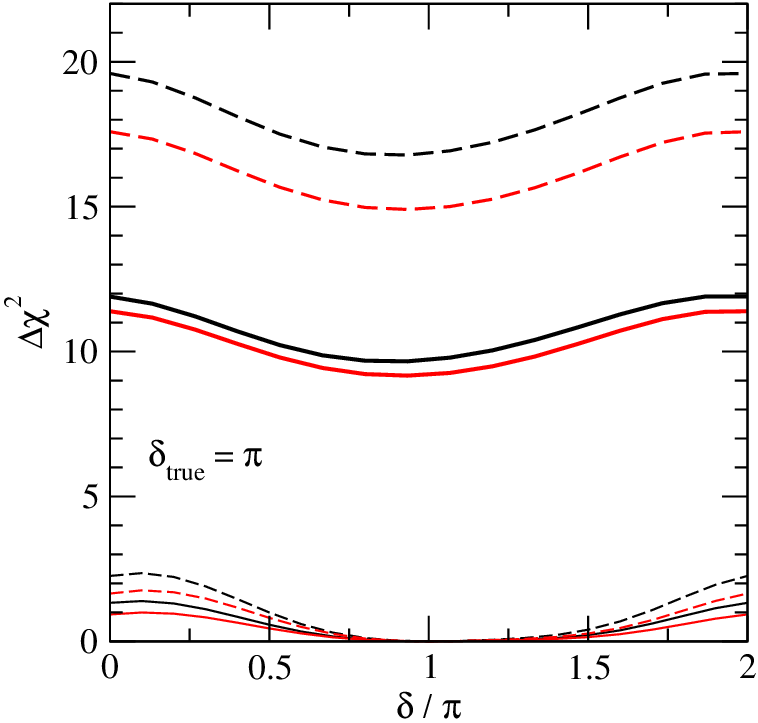}
\includegraphics[width=0.37\textwidth]{{chisq-del-1.5}.eps}
\end{center}
\mycaption{$\Delta\chi^2$ as a function of the CP phase $\delta$ for 3 years of PINGU data, for statistical errors only. We minimize with respect to $|\dmq|$, all other oscillation parameters are fixed. The four panels correspond to true values of $\delta = 0, \pi/2, \pi, 3\pi/2$. Thin curves with the minimum at $\Delta \chi^2 = 0$ correspond to the fit with the right mass ordering (normal), whereas the thick curves with non-zero minimum correspond to the wrong mass ordering (inverted). For black (red) curves the energy resolution is $\sigma_{E_\nu} = 0.2\%$ (2~GeV), for solid (dashed) curves the angular resolution is $\sigma_{\theta_\nu} = 1 \, (0.5) \times \sqrt{1 \, {\rm GeV} / E_\nu}$.}
\label{fig:PINGUdeltadependence}
\end{figure}

Let us now relax the assumption $\Delta m^2_{21} = 0$ and consider the
sensitivity when taking full three-flavour effects into account, including a
non-zero $\Delta m^2_{21}$ as well as the effect of the CP phase
$\delta$. In Fig.~\ref{fig:PINGUdeltadependence} we show the
$\Delta\chi^2$ of the wrong mass ordering as a function of $\delta$,
for various assumptions on the true CP phase. For this analysis only
statistical errors are assumed and all other oscillation parameters
except $|\dmq|$ are kept fixed. We observe that marginalizing over
$\delta$ has in impact of about 1 to 2 units in $\Delta\chi^2$ for the
exposure considered here (3 years). This corresponds roughly to a 10\%
effect on $\Delta\chi^2$. Note that this analysis is based on
statistics only, hence, the $\Delta\chi^2$ just scales linearly with
exposure. 

In passing let us comment on the possibility to constrain the CP phase
with PINGU. The thin curves in Fig.~\ref{fig:PINGUdeltadependence}
with the minimum at $\Delta\chi^2 = 0$ correspond to the fit with the
correct mass ordering. Those curves show that PINGU by itself will
have very poor sensitivity to the CP phase. One may expect that
systematic uncertainties will further reduce the sensitivity, even for
$\sim 10$~year exposures, see also~\cite{Winter:2013ema}.

\section{Daya Bay II analysis}
\label{sec:DB2}

The possibility to use a precision measurement of the $\bar\nu_e$
survival probability at a nuclear reactor to identify the neutrino
mass ordering~\cite{Petcov:2001sy} has been considered by a number of
authors~\cite{Schonert:2002ep, Choubey:2003qx, Learned:2006wy,
  Batygov:2008ku, Zhan:2008id, Zhan:2009rs, Ghoshal:2010wt,
  Ghoshal:2012ju, Qian:2012xh, Ciuffoli:2012bs, Ciuffoli:2012bp,
  Ge:2012wj, Ciuffoli:2013ep, Li:2013zyd}, boosted by the plans of the
Daya Bay and RENO collaborations for such an experiment.  For our
sensitivity calculations for Daya Bay II we will follow~\cite{DB2-YWang,
  DB2-WWang}. A 20~kt liquid scintillator detector is considered at
a distance of 58~km from the reactors with a total power of 36~GW. The
energy resolution is assumed to be $3\%\sqrt{1\,{\rm MeV}/E}$. We normalize our
number of events such that for an exposure of $\rm 20\, kt \times
36\,GW \times 6\, yr = 4320 \, kt \, GW \, yr$ we obtain $10^5$ events
\cite{DB2-YWang, DB2-WWang}.

In our analysis we assume that the neutrino source is point-like at a
distance of 58~km from the detector. We perform a $\chi^2$ analysis
using 350 bins for the energy spectrum. This number is chosen
sufficiently large such that bins are smaller (or of the order of) the
energy resolution. We take into account an overall normalization
uncertainty of 5\% and a linear energy scale uncertainty of
3\%. Uncertainties in the oscillation parameters $\sin^2\theta_{13}$
and $\sin^2\theta_{12}$ are included as pull parameters in the $\chi^2$
with $\sigma(\sin^2\theta_{13}) = 0.0023$ and
$\sigma(\sin^2\theta_{12}) = 0.012$. The $\chi^2$ analysis and its
interpretation is performed in complete analogy to the way described
in section~\ref{sec:chisq} for PINGU. With the above assumptions as well as $\sin^22\theta_{13} = 0.089$ and $\rm 4320 \, kt \, GW \, yr$ we find a
sensitivity to the mass ordering of $\Delta\chi^2 = 19$, which
compares reasonably well to the value $\Delta \chi^2 \approx 16$ found
in~\cite{ Li:2013zyd, DB2-YWang, DB2-WWang}. Our results are also in
reasonable agreement with~\cite{Qian:2012xh, Ge:2012wj} when we adopt
the same assumptions as there, however, we obtain significantly weaker
sensitivities as compared to~\cite{Ghoshal:2010wt, Ghoshal:2012ju}.

\begin{figure}
\begin{center}
\includegraphics[width=0.49\textwidth]{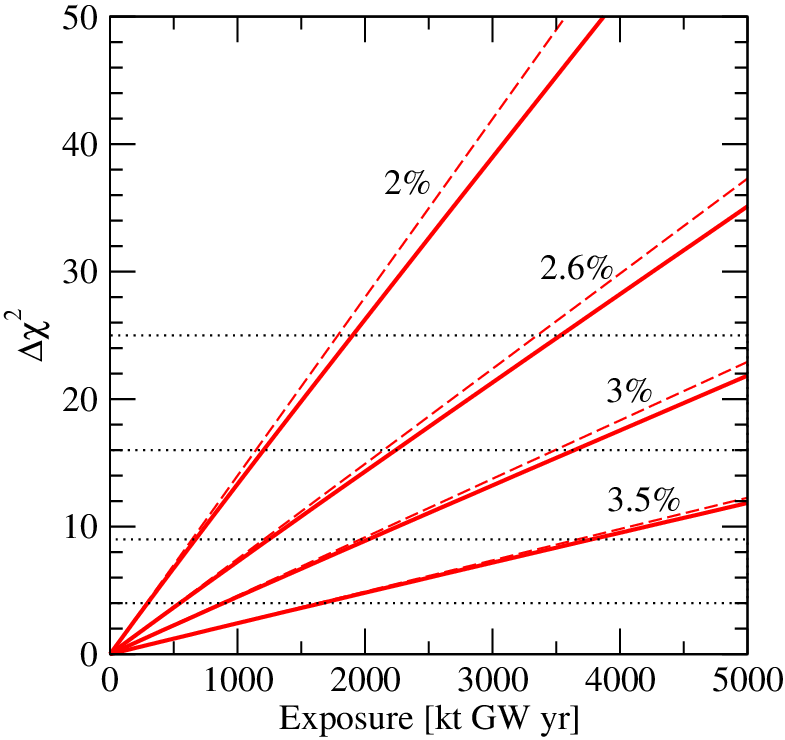}
\end{center}
\mycaption{$\Delta\chi^2$ of the wrong mass ordering for Daya Bay II as
  a function of the exposure for different assumptions on the energy
  resolution. The different set of curves correspond to energy
  resolutions of $\sigma_E/E = a \sqrt{1\,{\rm MeV}/E}$, with $a = 2\%, 2.6\%,
  3\%, 3.5\%$ as indicated in the plot. Dashed curves are for
  statistical errors only, solid curves include the uncertainty on
  normalization, linear energy scale, $\sin^2\theta_{13}$, and
  $\sin^2\theta_{12}$. We take $\sin^22\theta_{13} = 0.089$ and
  minimize with respect to $|\dmq|$. \label{fig:DB2-lumi}}
\end{figure}

In Fig.~\ref{fig:DB2-lumi} we show the Daya Bay II sensitivity to the
mass ordering as a function of the exposure, highlighting once more
the well-known importance of the energy resolution. We observe that
the systematical uncertainties considered here only play a sub-leading
role. We note that these results are essentially independent of
the assumed true ordering.

Let us point out that our analysis ignores some possible challenges of the
experiment, such as the smearing induced by the contributions from
reactor cores at slightly different baselines~\cite{Li:2013zyd}, the
background from more distant nuclear power plants, or the effect of a
non-linearity in the energy scale uncertainty
\cite{Qian:2012xh}. While such issues have to be addressed in the
actual analysis of such an experiment, our somewhat simplified
treatment suffices to illustrate the power of the atmospheric/reactor
combination.

\section{Combination of PINGU and Daya Bay II}
\label{sec:comb}

We now move to the main point of this work, the combination of data
from a high-statistics atmospheric and a medium-baseline
reactor experiment. For our combined analysis of PINGU and Daya Bay~II, we need to consider the full three flavor framework in order to
properly assess the combined sensitivity. This is due to the fact that
the effect we are exploiting is mainly based on the impact of $\Delta
m_{21}^2$ on the best fit of \dmq\ for the wrong ordering. It
is therefore necessary to take three flavour oscillations into account without
approximation in order to obtain reliable results. For computational
reasons we neglect the impact of systematic uncertainties in PINGU,
however we will comment on their impact later in this section.

\begin{figure}[t]
\begin{center}
\includegraphics[width=0.44\textwidth]{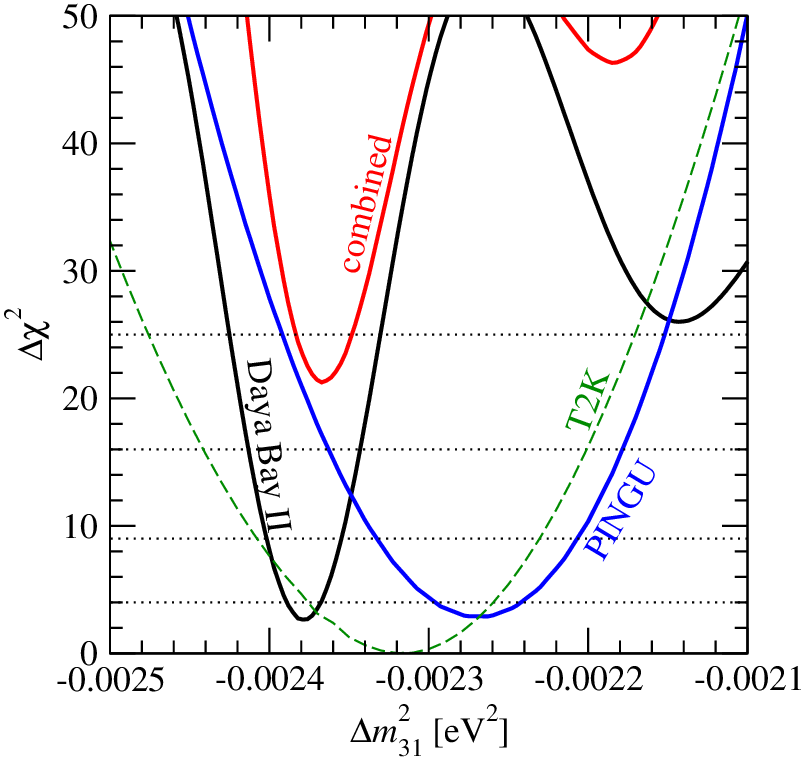}
\includegraphics[width=0.44\textwidth]{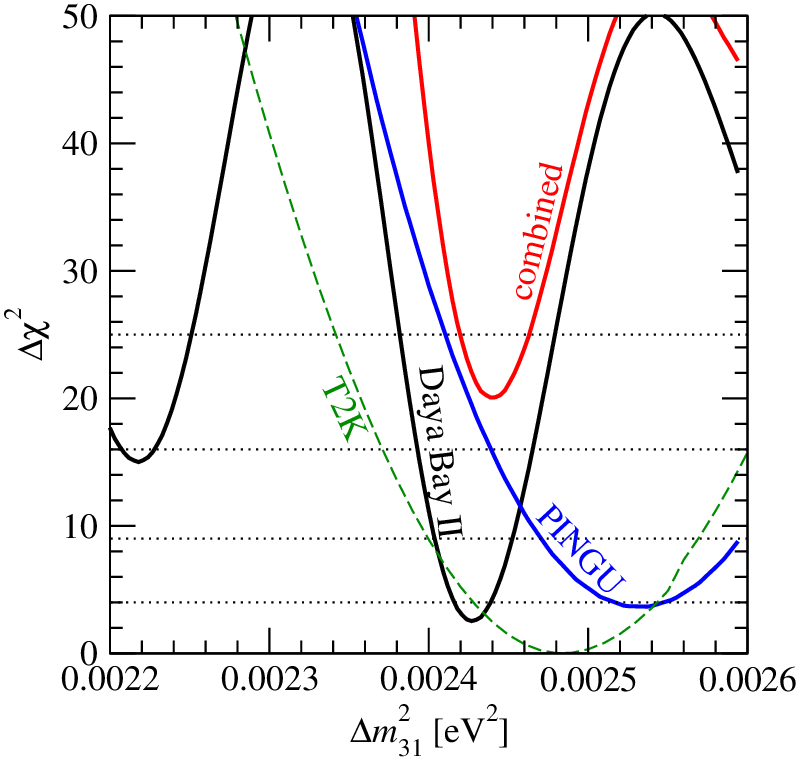}
\end{center}
\mycaption{$\Delta \chi^2$ as a function of $\dmq$ with the wrong sign
  for PINGU, Daya Bay II, and the combination. For PINGU we assume
  1~year of data with $\sigma_E = 2$~GeV and $\sigma_{\theta_\nu} = 
  \sqrt{1 \, {\rm GeV} / E_\nu}$, statistical errors only, and
  we minimize with respect to $\delta$ but keep all other oscillation
  parameters fixed. For Daya Bay~II we take an exposure of
  1000~kt~GW~yr and assume an energy resolution of $\sigma_E = 3.5\%
  \sqrt{1\,{\rm MeV}/E}$. The dashed curves corresponds to 5~years of
  neutrino data at 0.77~MW from T2K (not included in the ``combined''
  curve). We take the true values $|\dmq| = 2.4\times
  10^{-3}$~eV$^2$, $\sin^22\theta_{13} = 0.092$, $\sin^2\theta_{23} =
  0.5$, $\delta = 0$, $\Delta m^2_{21} = 7.59\cdot 10^{-5}~{\rm eV}^2$.
  For the left (right) panel the true mass ordering is normal (inverted).
\label{fig:sum}}
\end{figure}

The basic mechanism is illustrated in Fig.~\ref{fig:sum}. We show the
power of combining PINGU and Daya Bay~II results by plotting the
individual $\Delta\chi^2$ as well as their sum as a function of the wrong sign
\dmq. With the parameters chosen for this plot
neither of the experiments would have a sensitivity to the neutrino
mass ordering of more than two sigma. However, the $|\dmq|$ best fit
values would differ significantly. This implies that the overall best
fit occurs at a value of $|\dmq|$ which is not advantageous for either of
the experiments and therefore the sensitivity increases significantly,
as can be seen from the red curve, to between four and five sigma.

We can estimate the synergy effect in the following way: Let us denote
the individual $\Delta\chi^2$ functions as $\Delta\chi^2_i(x)$, with
$i = 1,2$, corresponding to PINGU and Daya Bay~II, respectively, and
$x$ denoting $|\dmq|$ in the wrong mass ordering. Close to the minimum
we can approximate the $\Delta\chi^2$ by a parabola and we write
\begin{equation}
  \Delta\chi^2_i(x) = \chi^2_{i,\text{min}} + \left(\frac{x - x_{0,i}}{\sigma_i}\right)^2 \,,
\end{equation}
where $x_{0,i}$ denotes the best fit value for $|\dmq|$ in the wrong
mass ordering and $\sigma_i$ the corresponding $1\sigma$ error for
experiment $i$. The minimum of the combined $\chi^2$ is then obtained as
\begin{equation}\label{eq:sumapprox}
  \Delta\chi^2_\text{comb} = \chi^2_{1,\text{min}} + \chi^2_{2,\text{min}} + 
  \frac{(x_{0,1} - x_{0,2})^2}{\sigma_1^2 + \sigma_2^2} \,.
\end{equation}
The first two terms correspond to the individual mass ordering
sensitivities of the two experiments, whereas the last term takes into
account the synergy effect from the mismatch of the different $\dmq$
best fit values. We observe that, if the best fit points differ by
more than the respective uncertainties summed in square, a relevant
synergy is obtained from the combination. Eq.~\ref{eq:sumapprox}
assumes that the parabolic shape of the individual $\chi^2$ functions
is still valid at the location of the combined best fit point, which
may not be true if it is located far away (in units of $\sigma_i$)
from one of the individual minima. In such a situation
eq.~\ref{eq:sumapprox} does not apply and the actual $\chi^2$ profiles
have to be used. Note also that eq.~\ref{eq:sumapprox} ignores
additional correlations due to the dependence on other oscillation
parameters (apart from $|\Delta m^2_{31}|$), which are, however,
expected to be small.

\begin{figure}[t]
\begin{center}
\includegraphics[width=0.48\textwidth]{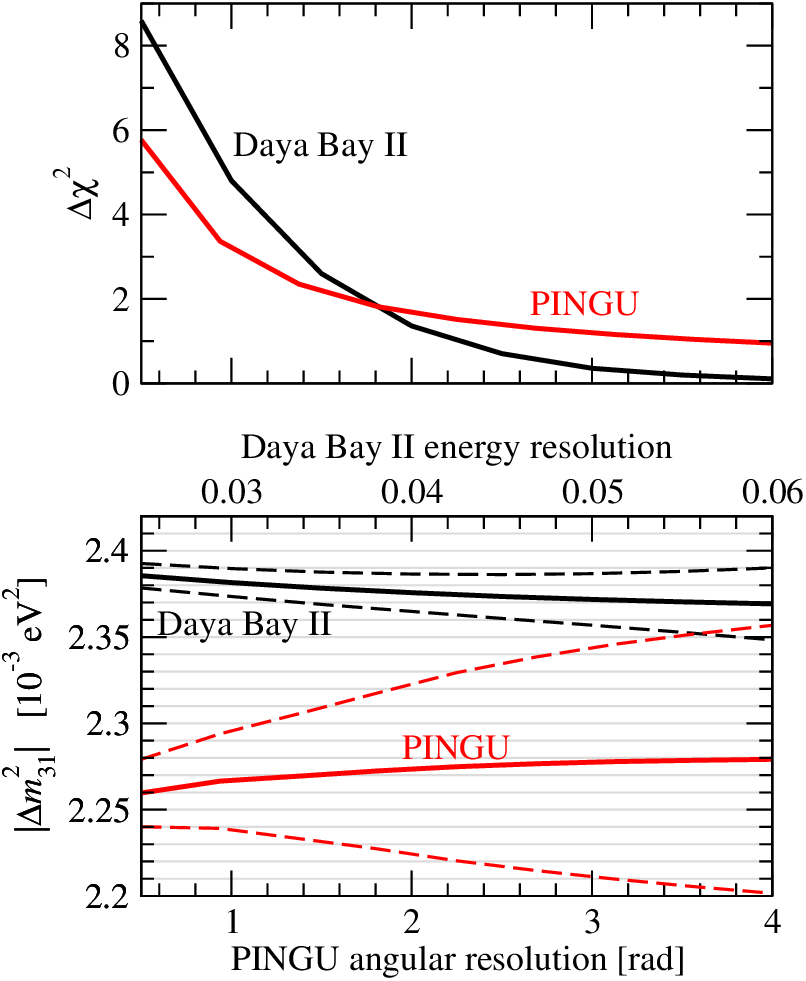}
\includegraphics[width=0.48\textwidth]{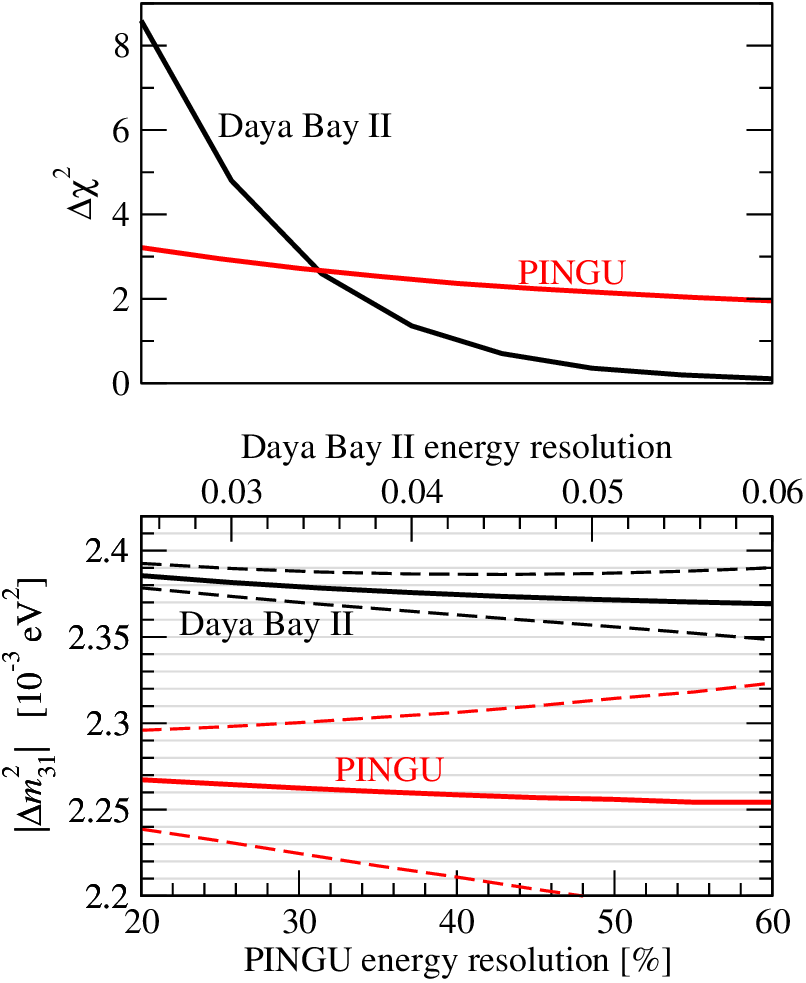}
\end{center}
\mycaption{Lower panels: the best fit (solid) and $1\sigma$ range
  (dashed) for $|\dmq|$ for the wrong mass ordering. For Daya Bay II
  we show it as a function of the energy resolution parameter $a$,
  where $\sigma_E/E_\nu = a\sqrt{1\,{\rm MeV}/E}$ (upper horizontal axis). The
  exposure is 1000~kt~GW~yr.  For PINGU we use the parametrization
  from eq.~\ref{eq:pingu-res}. In the left panel we take $A=0$,
  $B=0.2$, $C$ is shown on the lower horizontal axis; in the right
  panel we take $A=0$, $B$ is shown on the lower horizontal axis, $C =
  1$. The PINGU exposure is 1~year. In the upper panels we show
  the correspond $\Delta\chi^2$ for PINGU and Daya Bay~II alone. The
  true values are $\dmq = 2.4\cdot 10^{-3}\,{\rm eV}^2$ (normal
  ordering), $\sin^22\theta_{13} = 0.092$ and other parameters as in
  eq.~\ref{eq:true-values}. For PINGU we consider statistical errors
  only and we marginalize with respect to $\delta$, all other
  oscillation parameters are fixed.
\label{fig:resolutions}}
\end{figure}

In Fig.~\ref{fig:resolutions} we show how the best fit points for
$|\dmq|$ and its accuracy depend on experimental parameters for Daya
Bay~II and PINGU. We observe that the location of the minima are
relatively stable with respect to resolutions, and the difference
between the best fit points with the wrong sign of \dmq\ remains at
the level of $\gtrsim 0.1\cdot 10^{-3}$~eV$^2$. However, the accuracy
on $|\dmq|$ (especially from PINGU) is affected by resolutions. For
the chosen parameters an angular reconstruction in PINGU better than
about $2 \sqrt{1\,{\rm GeV}/E_\nu}$ is required to obtain a
sufficient accuracy on $\dmq$. In the upper panels we show the
individual sensitivities to the mass ordering. It is clear that the
synergy effect can be used to exclude the wrong mass ordering in a
regime where the individual experiments achieve only a poor rejection
power. This is particularly true for Daya Bay~II: while the
$\Delta\chi^2$ from Daya Bay~II alone is severely affected by the
energy resolution, the accuracy on $|\dmq|$ in the wrong ordering
remains excellent, see also \cite{Choubey:2003qx}. For instance even
for an energy resolution of 6\%, where the sensitivity to the mass
ordering completely disappears, an accuracy of $\sigma(|\dmq|) \approx
0.02\cdot 10^{-3}$~eV$^2$ is achieved, still much smaller than the
typical difference between the PINGU and Daya Bay~II best fit points,
which are of order $0.1\cdot 10^{-3}$~eV$^2$.

We note that for this plot only modest exposure times are assumed,
corresponding to 1~year of PINGU data and about 1.3~years reactor data for
a 20~kt detector at a 36~GW power plant. For the analysis used in
Fig.~\ref{fig:resolutions} the accuracy on $|\dmq|$ would scale with
the square-root of the exposure for Daya Bay~II as well as for PINGU.

We have checked that the accuracy on $|\dmq|$ as well as its best fit
value for the wrong ordering from PINGU depends very little on the
true value of $\theta_{23}$. Hence, the combined analysis with Daya
Bay~II remains unaffected by the value of $\theta_{23}$, whereas the
PINGU sensitivity to the mass ordering depends crucially on
$\theta_{23}$, as we have seen in Fig.~\ref{fig:octantPINGU}.

In Fig.~\ref{fig:sum} we show as dashed curves also the accuracy
obtainable by the T2K long-baseline experiment~\cite{Abe:2012gx}. We
use GLoBES~\cite{Huber:2004ka, Huber:2007ji} to simulate T2K and
assume 5 years of neutrino data at a beam power of 0.77~MW and we
consider only the $\nu_\mu$ disappearance channel, since we are
interested here in the obtainable accuracy on $|\dmq|$. We find a
minor synergy of Daya Bay~II also with T2K, although the effect is
much less significant compared to PINGU. We also observe that the
minimum of the wrong ordering appears for T2K at a slightly different
location than for PINGU, indicating that the matter effect plays an
important role in determining the best fit value of $|\dmq|$ with the
wrong sign. Let us comment also on a possible synergy of Daya Bay~II
with the INO atmospheric neutrino experiment~\cite{INO}. Using our
results from~\cite{Blennow:2012gj} we obtain an accuracy on $|\dmq|$
of $0.076 \,(0.054) \cdot 10^{-3}$~eV$^2$ for a 10~year exposure of a
50~kt (100~kt) detector, assuming the ``optimistic'' resolutions from
\cite{Blennow:2012gj}. While those accuracies are worse than the ones
we find for PINGU (depending on the resolutions) they are still
comparable to the difference in the $|\dmq|$ best fit points and at
some level a synergy effect may also emerge from the combination of
Daya Bay~II and INO. In~\cite{Ghosh:2012px} a slightly worse accuracy
of about $0.12 \cdot 10^{-3}$~eV$^2$ has been obtained for a 10~year
exposure of a 50~kt detector. A detailed numerical study of the
INO/Daya Bay~II combination is beyond the scope of this work.

Let us now comment on the possible impact of systematic uncertainties
on the PINGU result. This can be deduced from Fig.~\ref{fig:dm2}. This
figure has been obtained in the approximation $\Delta m^2_{21}=0$ and
therefore those results cannot be directly used for the PINGU/Daya
Bay~II comparison. However, we note that the best fit in the wrong
ordering occurs for values of $|\Delta m^2|$ smaller (larger) than the
simulated $2.4\cdot 10^{-3}$~eV$^2$ for a true normal (inverted)
ordering, although we used the approximation $\Delta m_{21}^2 =
0$. This shows that the physics involved here are fundamentally
different from the effect discussed
in~\cite{Nunokawa:2005nx,Minakata:2006gq} and rather depends on how
the matter effect comes into play. The final result will of course
depend on the matter effect as well as on $\Delta m^2_{21}$
effects. Indeed, both effects move the best fit in the same direction
and will therefore synergize to provide an even better sensitivity
when considering PINGU and Daya Bay~II together. This can be
appreciated by comparing the locations of the $\chi^2$ minima in
Figs.~\ref{fig:dm2} and \ref{fig:sum}, the latter showing much larger
deviations from the value assumed in the true mass ordering ($2.4\cdot
10^{-3}$~eV$^2$).

Nevertheless, we can use the results from Fig.~\ref{fig:dm2} to
estimate the impact of systematical errors.  We find that the
inclusion of systematical errors into the PINGU analysis does neither
change the location of the minimum in $|\dmq|$ nor the accuracy of its
determination. Instead, the main effect is a shift of the
$\Delta\chi^2$, modifying the mass ordering sensitivity of PINGU
alone. We have numerically verified that using reconstructed neutrino
parameters the value of $|\Delta m^2|$ at the minimum remains
unaffected up to the sub-percent level, and the accuracy remains
constant at a values of $\sigma(|\Delta m^2|) = 0.011\,(0.014)\cdot
10^{-3}$~eV$^2$ for resolutions according to eq.~\ref{eq:pingu-res} of
$A = 0 \, (2\, {\rm GeV})$, $B = 0.2 \, (0)$, $C = 0.5 \,
(1)$. For using muon reconstruction parameters the error is
between $0.018$ and $0.020 \cdot 10^{-3}$~eV$^2$ for the finite
resolutions and between $0.013$ and $0.014 \cdot 10^{-3}$~eV$^2$ for
perfect muon resolutions (these numbers correspond to a 3 year
exposure, as adopted in Fig.~\ref{fig:dm2}). Thus, the inclusion of
systematics into the PINGU analysis would not significantly affect the
outcome apart from shifting the PINGU and combined curves of
Fig.~\ref{fig:sum} down by an amount of at most the PINGU sensitivity
while the synergy effect of the PINGU/Daya Bay~II combination remains
unaffected.

\section{Conclusions}
\label{sec:conclusions}

In this work we pointed out a synergy for the determination of the
neutrino mass ordering between atmospheric neutrino data in low-energy
extensions of neutrino telescopes such as PINGU or ORCA and a
medium-baseline reactor neutrino experiment such as Daya Bay~II or
RENO50. Identifying the mass ordering is a rather challenging task
for those type of experiments and we may face the unfortunate
situation that each of them reaches only a poor sensitivity. One
reason for the difficulty is that it is possible to obtain a
reasonable fit within the wrong mass ordering by adjusting the value
of $|\dmq|$. However, the oscillation physics in the atmospheric and
reactor neutrino experiments are very different. In the first case a
complicated superposition of oscillation channels is observed and a
major role is played by the matter effect in the earth, while in the
second case tiny wiggles in the energy spectrum due to the $\bar\nu_e$
survival probability in vacuum are used. Hence, the values of $|\dmq|$
which may fake the wrong mass ordering are expected to be different in
the two type of experiments. Our numerical analysis shows that the
best fit values of $|\dmq|$ typically differ by about $0.1\cdot
10^{-3}$~eV$^2$. This is large compared to the rather impressive
accuracy with which $|\dmq|$ will be determined by those experiments,
which is typically in the range between $0.01$ and $0.02\cdot 10^{-3}$~eV$^2$.
Hence, it may be possible to exclude the wrong mass ordering due to
the mismatch of the best fit values for $\dmq$ in the two type of
experiments, even if each experiment on its own cannot.

We have performed some preliminary estimates on how the value and the
accuracy of $|\dmq|$ in the wrong ordering are affected by
experimental parameters such as energy resolutions, directional
reconstruction of atmospheric neutrinos, systematical uncertainties,
and exposure.  Our results indicate that for experimental parameters
similar to the ones discussed for the PINGU and ORCA proposals, as well as those for Daya
Bay~II and RENO50, the synergy effect in a combined analysis
will boost the mass ordering sensitivity significantly. The
experimental requirements are somewhat relaxed compared to those of the individual
sensitivities. Our results will need to be confirmed by more realistic
simulations once detailed reconstruction abilities and experimental
uncertainties of the respective experiments become available. 

While the sensitivity of atmospheric neutrinos to the mass ordering
strongly depends on the true value of $\theta_{23}$ (better
sensitivity for larger $\theta_{23}$) the synergy effect with the
reactor data does not depend on $\theta_{23}$. In this respect, the
sensitivity of the combined reactor and atmospheric neutrino analysis is
stable against the uncertainty introduced by the unknown true value of
$\theta_{23}$.

In conclusion, the combined analysis of atmospheric and reactor
neutrino experiments proposed here may be the only way to
identify the mass ordering if the individual experiments can only achieve
poor sensitivities, or the mass ordering may be identified
already after a much shorter running time of the
experiments. Certainly the comparison of the $|\dmq|$ measurement in
different experiments will provide an important cross check for any
mass ordering determination. Perhaps the combined sensitivity of ($i$)
the earth matter effect, ($ii$) subtle interference terms in the
vacuum $\bar\nu_e$ survival probability, and ($iii$) the comparison of
$|\dmq|$ measurements in different experiments may finally determine
the sign of \dmq\ beyond doubt.

\bigskip

{\bf Acknowledgment.} We thank Davide Franco and Anselmo Meregaglia
for intense exchanges in an early stage of this work and Alexei
Smirnov for discussions. We are grateful to Ken Clark for providing
information on PINGU. We thank Sandhya Choubey for pointing out an
important typo in an earlier version of this manuscript. This work was
supported by the G\"oran Gustafsson Foundation
(M.B.). T.S.\ acknowledges partial support from the European Union FP7
ITN INVISIBLES (Marie Curie Actions, PITN-GA-2011-289442).

\bigskip

{\bf Note added.} After the completion of this work the Daya Bay~II
project has been named JUNO (Jiangmen Underground Neutrino
Observatory), see e.g., \cite{Wang}.

\end{document}